\documentclass{article}
\usepackage{spconfa4,amsmath,amssymb,bm,mathtools,graphicx,cite,microtype,url,flushend}

\newcommand{\vect}[1]{\boldsymbol{#1}}
\newcommand{\mat}[1]{\boldsymbol{#1}}
\newcommand{\C}{\mathbb{C}}
\newcommand{\Hh}{\mathrm{H}}
\newcommand{\tr}{\operatorname{tr}}
\newcommand{\diag}{\operatorname{diag}}
\newcommand{\argmax}{\operatorname*{arg\,max}}
\newcommand{\1}{\mathbf{1}}
\title{cSTMM: A UNIFIED COMPLEX SPHERICAL STUDENT'S $t$ MIXTURE MODEL FOR DIRECTIONAL STATISTICS IN MASK-BASED BLIND SPEECH SEPARATION}
\name{Nobutaka Ito}
\address{Artificial Intelligence Research Center, AIST, Japan}

\begin{document}
\ninept
\maketitle

\begin{abstract}
Directional-statistics-based mask-based blind speech separation (BSS) clusters normalized time--frequency (TF) observations from $M$ microphones on the complex unit sphere, without relying on plane-wave or spherical-wave assumptions. Existing methods use separately defined angular mixture models, which makes the effect of density shape difficult to isolate. This paper proposes a complex spherical Student's $t$ mixture model (cSTMM) that connects the complex angular central Gaussian mixture model (cACGMM), complex Bingham mixture model (cBMM), and complex Watson mixture model (cWMM) through the degrees of freedom $\nu$ and eigenvalue constraints. We derive a latent-scale expectation-maximization (EM) framework with  an approximate M-step based on high-concentration approximation (HCA).
On noise-free LibriSpeech mixtures reverberated using measured room impulse responses (RIRs), the development-selected value $\nu^\ast=1$ outperformed the cACGMM-equivalent choice $\nu=M$ in all 18 test conditions, yielding a mean signal-to-distortion ratio improvement (SDRi) gain of $0.25\,\mathrm{dB}$. 
The model reduces to the cACGMM at $\nu=M$ and approaches the cBMM/cWMM in the large-$\nu$ limits.
\end{abstract}


\section{Introduction}
Time--frequency masks support BSS and speech enhancement by estimating source spectra directly or spatial covariance matrices for beamforming~\cite{araki2007,mandel2010,vuh10,sawada2010perm,yoshioka2015,heymann2015,ito2016cBMM,ito2016cACGMM,wang2018overview,boeddecker2018,raj2023gss,cornell2023chime7}. They may be predicted by supervised neural networks~\cite{wang2018overview,heymann2015} or estimated from each recording with spatial mixture models~\cite{araki2007,mandel2010,vuh10,sawada2010perm,ito2016cBMM,ito2016cACGMM,yoshioka2015,boeddecker2018,raj2023gss,cornell2023chime7}. The latter remain useful when representative training data or clean references are unavailable~\cite{boeddecker2018,raj2023gss,cornell2023chime7}.

Among spatial mixture models, 
directional-statistics-based mixture models cluster normalized multichannel observations on the complex unit sphere, without a plane-wave or spherical-wave model. The cWMM~\cite{vuh10} uses the complex Watson density~\cite{mardia1999} and ties dispersion across nonprincipal directions. The cBMM~\cite{ito2016cBMM} relaxes this constraint through the complex Bingham density~\cite{kent1994}. The cACGMM~\cite{ito2016cACGMM} instead uses the complex angular central Gaussian density~\cite{kent1997}. It outperformed the cWMM and the cBMM in an earlier comparison~\cite{ito2016cACGMM} and is used in guided source separation for the CHiME challenges~\cite{boeddecker2018,raj2023gss,cornell2023chime7}. Because these models are defined separately, however, they do not permit a controlled comparison of density shapes within a common parametric family.

We propose the cSTMM, which recovers the cACGMM at $\nu=M$ and, as $\nu\to\infty$, tends to the cBMM in its full-rank form and to the cWMM under tied nonprincipal eigenvalues. We derive latent-scale EM updates using HCA for the eigenvalues and compare the raw HCA update with a backtracked variant that accepts each proposal only when it increases the M-step objective. We then investigate the effect of $\nu$ in controlled multichannel BSS experiments by using the same mixtures and initialization across all $\nu$ values.

\section{Directional-statistics-based mask estimation}
Let $t$ and $f$ index time frames and frequency bins, respectively. Because the model is fitted independently at each frequency, we fix
$f$ and suppress its index hereafter. Let $\vect{y}_{t}\in\C^M$ denote the short-time Fourier transform (STFT) vector observed by $M$ microphones  at the fixed frequency, and let $N$ be the number of sources. We normalize each valid
observation as
\begin{equation}
\vect{z}_{t}=\frac{\vect{y}_{t}}{\|\vect{y}_{t}\|_2},
\qquad
\mathcal{T}=\{t:\|\vect{y}_{t}\|_2\geq\varepsilon\},
\label{eq:dirfeat}
\end{equation}
where $\mathcal{T}$ is the set of valid frames retained for model fitting. If source $n$ dominates a TF bin, the sparse approximation $\vect{y}_{t}\approx\vect{h}^{(n)}s^{(n)}_{t}$ implies that $\vect{z}_{t}$ primarily depends on its transfer-function vector $\vect{h}^{(n)}$. Let $d_{t}\in\{1,\ldots,N\}$ denote the dominant-source label. We model
the normalized observations at the fixed frequency by
\begin{equation}
p(\vect{z}_{t};\Theta)
=\sum_{n=1}^{N}w^{(n)}p(\vect{z}_{t}\mid d_{t}=n;\Theta),
\label{eq:mixture}
\end{equation}
where $w^{(n)}\geq0$, $\sum_n w^{(n)}=1$. For a fixed value of $\nu$, $\Theta$ denotes the parameters estimated at the current frequency.
We estimate
${\Theta}$ by maximizing
$\sum_{t\in\mathcal{T}}\log p(\vect{z}_{t};\Theta)$. The posterior
responsibility of component $n$ is
\begin{equation}
\gamma_{t}^{(n)}
=\frac{w^{(n)}p(\vect{z}_{t}\mid d_{t}=n;\Theta)}
{\sum_{n'=1}^{N}w^{(n')}p(\vect{z}_{t}\mid d_{t}=n';\Theta)},
\label{eq:mask_generic}
\end{equation}
which serves as a soft mask.  Because the models are fitted independently at different
frequencies, their component labels are arbitrary across frequency.
We therefore align the component permutations after all
frequency-wise models have been fitted~\cite{sawada2010perm}. The
aligned masks are then either applied to a reference-microphone STFT
to estimate the source spectra directly or used as weights to estimate
source-wise covariance matrices for beamforming.

\section{Proposed method}
\subsection{Full cSTMM and its limiting cases}
For source $n$ at the fixed frequency, the full cSTMM has a weight $w^{(n)}$ and a Hermitian parameter matrix $\mat{A}^{(n)}$. For degrees of freedom $\nu>0$, its component density with respect to the normalized uniform measure on the complex unit sphere is~\cite{kato2004}
\begin{equation}
p(\vect{z};\mat{A},\nu)
=C(\mat{A},\nu)
\left(1-\frac{2}{\nu}\vect{z}^{\Hh}\mat{A}\vect{z}\right)^{-a_\nu},
\quad a_\nu=\frac{\nu+M}{2},
\label{eq:cst}
\end{equation}
where $\|\vect{z}\|_2=1$ and $\lambda_{\max}(\mat{A})<\nu/2$. For a Hermitian $\mat{B}$ with $b=\lambda_{\max}(\mat{B})<\nu/2$, define
\begin{equation}
\operatorname{can}_{\nu}(\mat{B})
=\frac{\mat{B}-b\mat{I}}{1-2b/\nu}.
\label{eq:canonicalization}
\end{equation}
Because
\begin{equation}
1-\frac{2}{\nu}\vect{z}^{\Hh}\operatorname{can}_{\nu}(\mat{B})\vect{z}
=\frac{1-(2/\nu)\vect{z}^{\Hh}\mat{B}\vect{z}}{1-2b/\nu},
\label{eq:canonical_equiv}
\end{equation}
$\mat{B}$ and $\operatorname{can}_{\nu}(\mat{B})$ differ only by a $\vect{z}$-independent kernel factor and therefore define the same normalized density. The zero-eigenvalue eigenspace of the canonical matrix is the principal eigenspace of $\mat{B}$. We henceforth parameterize each component by
this canonical representative, so that
$\mat{A}\preceq\mat{O}$ and $\lambda_{\max}(\mat{A})=0$.

Let
$\mat{A}
=\mat{U}\operatorname{diag}(0,\lambda_2,\ldots,\lambda_M)
\mat{U}^{\Hh}$ be an eigendecomposition
and $\widetilde{\vect{z}}=\mat{U}^{\Hh}\vect{z}$.
When $\bm{z}$ is uniformly distributed on the complex unit sphere,
$\bigl(
|\widetilde z_1|^2,\ldots,|\widetilde z_M|^2
\bigr)
\sim\operatorname{Dirichlet}(1,\ldots,1)$.
Set $q_j=|\widetilde z_j|^2$ and
$\vect{q}=(q_2,\ldots,q_M)^{\mathsf T}$, so that
$q_1=1-\sum_{j=2}^{M}q_j$. The inverse normalizer is
\begin{equation}
C(\mat{A},\nu)^{-1}
=
\Gamma(M)
\int_{\Delta_{M-1}}
\left(
1-\frac{2}{\nu}\sum_{j=2}^{M}\lambda_jq_j
\right)^{-a_\nu}
d\vect{q},
\label{eq:simplex_normalizer}
\end{equation}
where
\begin{equation}
\Delta_{M-1}
=
\left\{
\vect{q}\in\mathbb{R}^{M-1}_{\geq 0}\mid
\sum_{j=2}^{M}q_j\leq1
\right\}.
\end{equation}
We evaluate this integral numerically.

For fixed $\mat{A}$, as $\nu\to\infty$, the density converges to the
complex Bingham density because its kernel tends to
$\exp(\vect{z}^{\Hh}\mat{A}\vect{z})$.
At $\nu=M$, setting $\mat{P}=\mat{I}-2\mat{A}/M$ gives
$p(\vect{z})\propto(\vect{z}^{\Hh}\mat{P}\vect{z})^{-M}$, the complex
angular central Gaussian density, with
$\lambda_{\min}(\mat{P})=1$ fixing its scale.

\subsection{Rank-one constraint and the cWMM limit}
Let $\mat{B}=\rho\vect{a}\vect{a}^{\Hh}$, where $0\leq\rho<\nu/2$ and $\|\vect{a}\|_2=1$. Its canonical form is
\begin{equation}
\mat{A}=-\kappa(\mat{I}-\vect{a}\vect{a}^{\Hh}),
\qquad
\kappa=\frac{\rho}{1-2\rho/\nu}.
\label{eq:rankone_A}
\end{equation}
For $\rho>0$, $\vect{a}$ is the principal eigenvector, with eigenvalue zero. The remaining $M-1$ eigenvalues equal $-\kappa$. The density becomes
\begin{equation}
p(\vect{z};\vect{a},\kappa,\nu)\propto
\left[1+\frac{2\kappa}{\nu}
\left(1-|\vect{a}^{\Hh}\vect{z}|^2\right)\right]^{-a_\nu}.
\label{eq:cst-r1}
\end{equation}

The inverse normalizer of \eqref{eq:cst-r1} is obtained from the
simplex integral for the full cSTMM in
\eqref{eq:simplex_normalizer} by setting
$\lambda_2=\cdots=\lambda_M=-\kappa$.
Choosing an eigendecomposition whose first column of $\mat{U}$ is
$\vect{a}$ gives
\begin{equation}
q_1=|\vect{a}^{\Hh}\vect{z}|^2,
\qquad
S=\sum_{j=2}^{M}q_j
=1-|\vect{a}^{\Hh}\vect{z}|^2.
\end{equation}
By the aggregation property of the Dirichlet distribution,
$S\sim\operatorname{Beta}(M-1,1)$. Hence,
\begin{equation}
C(\kappa,\nu)^{-1}
=
(M-1)
\int_0^1
s^{M-2}
\left(
1+\frac{2\kappa}{\nu}s
\right)^{-a_\nu}
ds,
\label{eq:rankone_normalizer}
\end{equation}
and we evaluate this one-dimensional integral numerically.

For fixed $\vect{a}$ and $\kappa$, as $\nu\to\infty$, the kernel in
\eqref{eq:cst-r1} converges uniformly to
\begin{equation}
\exp\!\left[
-\kappa
\left(
1-|\vect{a}^{\Hh}\vect{z}|^2
\right)
\right]
=
\exp(-\kappa)
\exp\!\left(
\kappa|\vect{a}^{\Hh}\vect{z}|^2
\right).
\end{equation}
The factor $\exp(-\kappa)$ cancels upon normalization, so the
normalized density converges to the complex Watson density.

\subsection{Latent-scale EM algorithm}
This subsection first derives the full-cSTMM updates and then gives the rank-1 counterparts. At the fixed frequency, the objective is $\mathcal{L}(\Theta)=\sum_{t\in\mathcal{T}}\log p(\vect{z}_{t};\Theta)$. Define
$r(\vect{z},\mat{A})=1-(2/\nu)\vect{z}^{\Hh}\mat{A}\vect{z}$. Introduce a positive latent scale variable $u$. The cST component
density in \eqref{eq:cst} is the marginal of the joint density
\begin{equation}
p(\vect{z},u;\mat{A},\nu)
=\frac{C(\mat{A},\nu)}{\Gamma(a_\nu)}
 u^{a_\nu-1}\exp[-u\,r(\vect{z},\mat{A})],
\quad u>0,
\label{eq:joint_aug}
\end{equation}
which integrates to~\eqref{eq:cst}. 

At the current parameters $\Theta^{\mathrm{old}}$, let
\begin{align}
r_{t}^{(n)}
&=1-\frac{2}{\nu}\vect{z}_{t}^{\Hh}
\mat{A}^{(n),\mathrm{old}}\vect{z}_{t},\label{eq:r}\\
\ell_{t}^{(n)}
&=w^{(n),\mathrm{old}}C(\mat{A}^{(n),\mathrm{old}},\nu)
(r_{t}^{(n)})^{-a_\nu}.\label{eq:ell}
\end{align}
The E-step gives
\begin{align}
\gamma_{t}^{(n)}
&=\frac{\ell_{t}^{(n)}}{\sum_{n'=1}^{N}\ell_{t}^{(n')}},\label{eq:gamma_tight}\\
u_{t}\mid\vect{z}_{t},d_{t}=n
&\sim\operatorname{Gamma}(a_\nu,r_{t}^{(n)}),\label{eq:u_posterior}\\
\bar u_{t}^{(n)}
&\coloneqq\mathbb{E}[u_{t}\mid\vect{z}_{t},d_{t}=n;\Theta^{\mathrm{old}}]\\
&=\frac{a_\nu}{r_{t}^{(n)}}.\label{eq:u_mean}
\end{align}
Here, $\operatorname{Gamma}(\alpha,\beta)$ denotes the shape--rate
distribution with density
$\beta^\alpha u^{\alpha-1}e^{-\beta u}/\Gamma(\alpha)$ for $u>0$.
Up to terms independent of $\Theta$, the EM auxiliary function is
\begin{align}
Q(\Theta\mid\Theta^{\mathrm{old}})
&=\sum_{t\in\mathcal{T}}\sum_{n=1}^{N}\gamma_{t}^{(n)}
\left[\log w^{(n)}+\log C(\mat{A}^{(n)},\nu)\right.\notag\\
&\quad\left.+\frac{2}{\nu}\bar u_{t}^{(n)}
\vect{z}_{t}^{\Hh}\mat{A}^{(n)}\vect{z}_{t}\right].
\label{eq:em_Q}
\end{align}
Thus
\begin{equation}
w^{(n)}\leftarrow\frac{1}{|\mathcal{T}|}
\sum_{t\in\mathcal{T}}\gamma_{t}^{(n)}.
\label{eq:w_update}
\end{equation}
With
\begin{equation}
G^{(n)}=\sum_{t\in\mathcal{T}}\gamma_{t}^{(n)},
\qquad
\mat{S}^{(n)}=\frac{2}{\nu}
\sum_{t\in\mathcal{T}}\gamma_{t}^{(n)}\bar u_{t}^{(n)}
\vect{z}_{t}\vect{z}_{t}^{\Hh},
\label{eq:Slatent}
\end{equation}
the matrix-dependent objective is
\begin{equation}
G^{(n)}\log C(\mat{A}^{(n)},\nu)
+\tr\!\left(\mat{A}^{(n)}\mat{S}^{(n)}\right).
\label{eq:em_Aterms}
\end{equation}
Let the eigendecompositions be
\begin{align}
\mat{A}^{(n)}
&=\mat{U}^{(n)}\diag(\lambda_{1}^{(n)},\ldots,\lambda_{M}^{(n)})
\mat{U}^{(n)\Hh},\label{eq:Aevd}\\
\mat{S}^{(n)}
&=\mat{V}^{(n)}\diag(\sigma_{1}^{(n)},\ldots,\sigma_{M}^{(n)})
\mat{V}^{(n)\Hh},\label{eq:Sevd}
\end{align}
where $\mat{U}^{(n)}$ and $\mat{V}^{(n)}$ are unitary,
$0=\lambda_{1}^{(n)}\geq\cdots\geq\lambda_{M}^{(n)}$, and
$\sigma_{1}^{(n)}\geq\cdots\geq\sigma_{M}^{(n)}\geq0$.
Von Neumann's trace inequality aligns the ordered eigenvectors, so
$\mat{U}^{(n)}\leftarrow\mat{V}^{(n)}$. In particular, the principal eigenvector of $\mat{A}^{(n)}$ becomes that of $\mat{S}^{(n)}$.

To derive the HCA update for the nonprincipal eigenvalues, we
suppress the component index $(n)$ and work in the eigenbasis
$\mat{U}$ of $\mat{A}$. Let
$\widetilde{\vect{z}}=\mat{U}^{\Hh}\vect{z}$,
$\xi_j=\widetilde z_j/\widetilde z_1$ for
$j=2,\ldots,M$, and
$\vect{\xi}=(\xi_2,\ldots,\xi_M)^{\mathsf T}$.
Under the normalized uniform measure on the complex unit sphere,
$\widetilde{\vect{z}}$ is also uniformly distributed by unitary
invariance.  $\vect{\xi}$ is well
defined almost surely and follows the standard complex multivariate
Cauchy distribution on $\C^{M-1}$~\cite{KatoMcCullagh2014}.
Consequently,
\begin{align}
C(\mat{A},\nu)^{-1}
&=
\frac{\Gamma(M)}{\pi^{M-1}}
\int_{\C^{M-1}}
\left(
1-\frac{2}{\nu}
\frac{\sum_{j=2}^{M}\lambda_j|\xi_j|^2}
{1+\|\vect{\xi}\|_2^2}
\right)^{-a_\nu}
\notag\\
&\quad\times
\frac{d\vect{\xi}}
{\left(1+\|\vect{\xi}\|_2^2\right)^M}.
\label{eq:cst_xi_integral}
\end{align}
The HCA about $\vect{\xi}=\vect{0}$ replaces
$1+\|\vect{\xi}\|_2^2$ by one in both occurrences in
\eqref{eq:cst_xi_integral}. Assuming $\lambda_j<0$ for $j=2,\ldots,M$, changing variables as
$\eta_j=(-\lambda_j)^{1/2}\xi_j$, and restoring the component index
gives
\begin{equation}
\log C(\mat{A}^{(n)},\nu)
\approx
\sum_{j=2}^{M}
\log\!\left(-\lambda_j^{(n)}\right)
+\mathrm{const.}
\label{eq:hca_logC}
\end{equation}
Although the exact integral in \eqref{eq:cst_xi_integral} is
finite for every $\nu>0$, the integrand obtained after the HCA
replacement is integrable over $\mathbb{C}^{M-1}$ only for
$\nu>M-2$. 
Substituting \eqref{eq:hca_logC} into
\eqref{eq:em_Aterms} and assuming $\sigma^{(n)}_j>0$ for $j=2,\dots,M$ gives
\begin{equation}
\lambda_j^{(n),\mathrm{HCA}}
=
-\frac{G^{(n)}}{\sigma_j^{(n)}},
\qquad j=2,\ldots,M.
\label{eq:hca_prop}
\end{equation}

The rank-1 model uses the same $\gamma_{t}^{(n)}$, $\bar u_{t}^{(n)}$, and $w^{(n)}$ updates. Its HCA surrogate is
\begin{align}
\widetilde Q^{(n)}
&=G^{(n)}(M-1)\log\kappa^{(n)}\notag\\
&\quad-\kappa^{(n)}
\left[\tr(\mat{S}^{(n)})-
\vect{a}^{(n)\Hh}\mat{S}^{(n)}\vect{a}^{(n)}\right].
\label{eq:rankone_surrogate}
\end{align}
Hence $\vect{a}^{(n)}$ is a principal eigenvector of $\mat{S}^{(n)}$ and
\begin{equation}
\kappa^{(n),\mathrm{HCA}}
=\frac{G^{(n)}(M-1)}
{\tr(\mat{S}^{(n)})-\sigma_{1}^{(n)}}
=\frac{G^{(n)}(M-1)}{\sum_{j=2}^{M}\sigma_{j}^{(n)}}.
\label{eq:kappa_update}
\end{equation}

The closed-form HCA updates in
\eqref{eq:hca_prop} and \eqref{eq:kappa_update} generalize the
eigenvalue update for the complex Watson distribution
in~\cite{mardia1999} to the full and rank-1 cSTMMs,
respectively. We evaluate these expressions as HCA proposals
throughout the tested range, including $\nu\leq M-2$, both in raw
form and with objective-checked backtracking.

\begin{figure*}[t]
\centering
\includegraphics[width=0.75\textwidth,trim=7 10 9 10,clip]{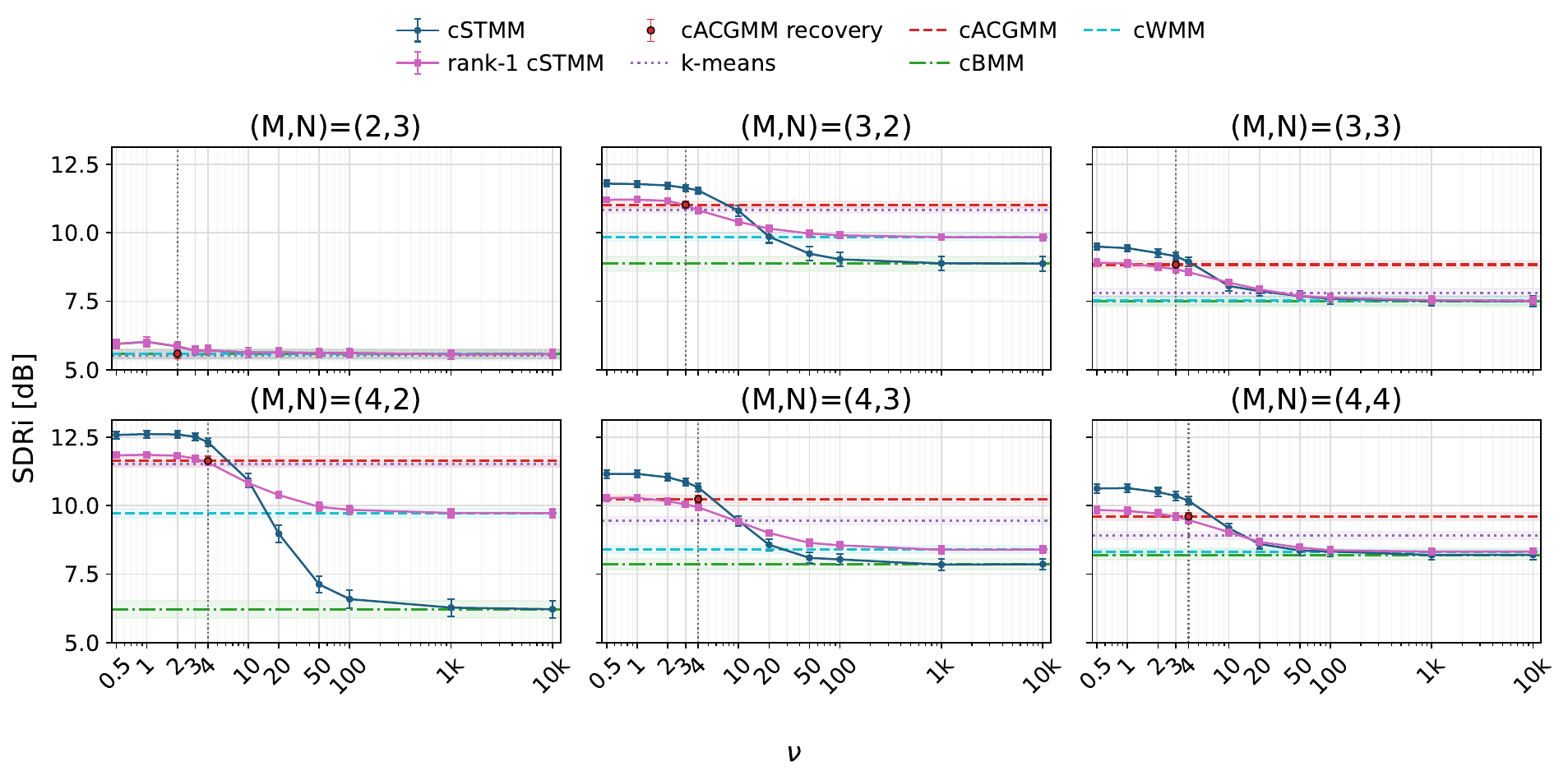}
\caption{Development-set SDRi versus $\nu$ at reverberation time $\mathrm{RT}_{60}=610\,\mathrm{ms}$ (64 mixtures per panel, matched initialization). The vertical dotted line marks $\nu=M$. The isolated red markers are denoted ``cACGMM recovery'' in the figure. Horizontal reference lines show k-means, the cACGMM, the cBMM, and the cWMM\@.}
\label{fig:sweep}
\end{figure*}

\begin{table}[t]
\centering
\setlength{\tabcolsep}{1.1pt}
\renewcommand{\arraystretch}{0.90}
\caption{Mean test SDRi (dB) for the $\nu^\ast=1$ and $\nu=M$ settings,
evaluated on the same 256 test mixtures per condition. Higher SDRi values are
better, and $\mathrm{RT}_{60}$ denotes the reverberation time.
$\Delta$ is the mean SDRi difference,
$\mathrm{SDRi}_{\nu^\ast=1}-\mathrm{SDRi}_{\nu=M}$, and SE is its
standard error. $p_{\mathrm{Holm}}$ is the $p$-value from the
Wilcoxon signed-rank test, adjusted by Holm's method across the
18 test conditions.}\vspace{2mm}
\label{tab:conditionwise}
\begin{tabular}{crrrrrrr}
\hline
$(M,N)$ & $\mathrm{RT}_{60}$ & $\nu^\ast$ & $\nu=M$ & $\Delta$ & SE & $p_{\mathrm{Holm}}$ \\
 & [ms] & [dB] & [dB] & [dB] & [dB] & & \\
\hline
$(2,3)$ & 160 & 10.640 & 10.580 & 0.060 & 0.021 & $6.4{\times}10^{-8}$ \\
$(2,3)$ & 360 & 8.815 & 8.790 & 0.025 & 0.027 & $6.4{\times}10^{-6}$ \\
$(2,3)$ & 610 & 6.259 & 6.210 & 0.049 & 0.042 & 0.001  \\
$(3,2)$ & 160 & 14.092 & 13.835 & 0.258 & 0.017 & $5.5{\times}10^{-34}$  \\
$(3,2)$ & 360 & 13.466 & 13.293 & 0.172 & 0.010 & $1.7{\times}10^{-37}$  \\
$(3,2)$ & 610 & 11.796 & 11.659 & 0.137 & 0.013 & $2.1{\times}10^{-29}$  \\
$(3,3)$ & 160 & 12.530 & 12.431 & 0.099 & 0.022 & $4.0{\times}10^{-9}$  \\
$(3,3)$ & 360 & 11.422 & 11.237 & 0.185 & 0.021 & $7.7{\times}10^{-18}$  \\
$(3,3)$ & 610 & 9.268 & 9.064 & 0.204 & 0.040 & $3.6{\times}10^{-13}$  \\
$(4,2)$ & 160 & 13.920 & 13.544 & 0.377 & 0.022 & $4.1{\times}10^{-36}$  \\
$(4,2)$ & 360 & 13.578 & 13.346 & 0.233 & 0.013 & $1.8{\times}10^{-35}$ \\
$(4,2)$ & 610 & 12.584 & 12.393 & 0.191 & 0.016 & $2.7{\times}10^{-28}$ \\
$(4,3)$ & 160 & 12.611 & 12.258 & 0.353 & 0.026 & $3.2{\times}10^{-30}$ \\
$(4,3)$ & 360 & 12.470 & 12.119 & 0.351 & 0.030 & $2.1{\times}10^{-28}$  \\
$(4,3)$ & 610 & 11.039 & 10.506 & 0.533 & 0.032 & $4.0{\times}10^{-35}$  \\
$(4,4)$ & 160 & 13.041 & 12.641 & 0.400 & 0.022 & $4.0{\times}10^{-35}$  \\
$(4,4)$ & 360 & 12.241 & 11.881 & 0.359 & 0.029 & $1.3{\times}10^{-25}$  \\
$(4,4)$ & 610 & 10.658 & 10.140 & 0.518 & 0.044 & $1.1{\times}10^{-27}$  \\
\hline
\end{tabular}
\end{table}

\section{Experimental evaluation}
\label{sec:experiments}
\subsection{Experimental conditions}
\textbf{Data.}
We formed 8-s development and test mixtures by convolving
16-kHz LibriSpeech \texttt{dev-clean} and \texttt{test-clean}
speech~\cite{panayotov2015}, respectively, with RIRs from MIRD~\cite{hadad2014mird}.
The speaker sets were disjoint. Each mixture used $N$ sampled speakers whose utterances were randomly permuted and concatenated. We used the MIRD \texttt{8-8-8-8-8-8-8} array RIRs at $1\,\mathrm{m}$. Array broadside was defined as $0^\circ$, with positive angles measured counterclockwise. The source directions were $60^\circ,-15^\circ,30^\circ$, and $-45^\circ$, and the first $N$ directions were used. For $M=2,3,4$, zero-based channel indices were $(3,4)$, $(3,4,5)$, and $(2,3,4,5)$. Source images were summed without additive noise, and mixture indices and random seeds were reused across conditions. 
We evaluated six $(M,N)$ settings, $(2,3)$, $(3,2)$, $(3,3)$,
$(4,2)$, $(4,3)$, and $(4,4)$, at three reverberation times,
$\mathrm{RT}_{60}=160$, 360, and $610\,\mathrm{ms}$, yielding
18 test conditions.

\textbf{Preprocessing and initialization.}
We used an STFT with a 2048-sample Hann window and a 512-sample hop. At each frequency, observations satisfying $\|\vect{y}_{t}\|_2\geq10^{-6}$ were unit-normalized, prewhitened, and renormalized as in~\cite{sawada2010perm}, with covariance eigenvalues floored at $10^{-8}$. Complex spherical k-means provided the initial hard clustering. In k-means run $r$, $N$ frames sampled without replacement from $\mathcal{T}$ initialized the $N$ centroids $\{\vect{a}^{(n,r)}\}_{n=1}^{N}$. Each run alternated data assignment and centroid updates until no assignment changed or 20 iterations elapsed. Data assignment used
\begin{equation}
d_{t}^{(r)}=\argmax_n
|\vect{a}^{(n,r)\Hh}\vect{z}_{t}|.
\label{eq:kmeans_assignment}
\end{equation}
The centroid update sets $\vect{a}^{(n,r)}$ to a unit-norm
principal eigenvector of
\begin{equation}
\mat{R}^{(n,r)}=
\frac{\sum_{t\in\mathcal{T}}\1\{d_{t}^{(r)}=n\}
\vect{z}_{t}\vect{z}_{t}^{\Hh}}
{\sum_{t\in\mathcal{T}}\1\{d_{t}^{(r)}=n\}}.
\label{eq:kmeans_centroid}
\end{equation}
We chose
$r^\ast=\argmax_r\sum_{t\in\mathcal{T}}
|\vect{a}^{(d_{t}^{(r)},r)\Hh}\vect{z}_{t}|^2$.
Its assignments initialized $\gamma_{t}^{(n)}=\1\{d_{t}^{(r^\ast)}=n\}$, and the weights $w^{(n)}$ were initialized using (22). $\mat{R}^{(n,r^\ast)}$ initialized
$\mat{A}^{(n)}=\operatorname{can}_{\nu}[(\nu/4)\mat{R}^{(n,r^\ast)}]$.

\textbf{$\nu$ selection and EM fitting.}
For each $(M, N)$ setting, candidate values of $\nu$ spanning
$0.5$ to $10^4$ were examined using 64 development mixtures at
$\mathrm{RT}_{60}=610\,\mathrm{ms}$. Based on aggregate
development-set SDRi across the six settings, the common value
$\nu^\ast=1$ was selected for the test-set evaluation. For each paired comparison, the $\nu=1$ and $\nu=M$ fits were
applied to the same mixture and initialized with the same
responsibilities $\gamma_t^{(n)}$. Five fixed-responsibility M-steps were used for warm-starting;
the EM iteration limit was 20. Let $\mathcal{L}^{(i)}$ denote the log-likelihood value used for
convergence monitoring at iteration $i$. Fitting stopped when the
relative change in $\mathcal{L}^{(i)}$ was below $10^{-6}$ for two
consecutive iterations. Unless otherwise stated, we report results obtained using the raw HCA proposal. Objective-checked backtracking was evaluated separately on the diagnostic subset described below.
Post-EM permutation alignment used the \texttt{DHTVPermutationAlignment} class in \texttt{pb\_bss},\footnote{\url{https://github.com/fgnt/pb_bss}} using an STFT size of 2048, cosine similarity, and greedy assignment.

\subsection{Main separation result}
Table~\ref{tab:conditionwise} compares $\nu^\ast=1$ with cACGMM-equivalent $\nu=M$. The gain reached $0.533\,\mathrm{dB}$ for $(M,N)=(4,3)$ at $\mathrm{RT}_{60}=610\,\mathrm{ms}$ but was 0.060, 0.025, and $0.049\,\mathrm{dB}$ in the three $(2,3)$ conditions. Because $(2,3)$ is both the only two-microphone and the only underdetermined setting, these factors cannot be separated here.

For an oracle diagnostic
using the reference source images, a valid
TF bin was labeled ambiguous when the largest source-image power
was less than 0.7 times the summed source-image power. For each mixture, the source-assignment entropy $-\sum_n\gamma^{(n)}_t\log\gamma^{(n)}_t$ and maximum assignment probability $\max_n\gamma^{(n)}_t$ were averaged over all ambiguous bins in time and frequency. Relative to $\nu=M$, $\nu=1$ increased mean entropy and decreased the mean maximum probability for all $18\times256=4608$ paired method evaluations. For $N=2$, 3, and 4, respectively, the mean entropy changed from
0.336 to 0.434, from 0.556 to 0.689, and from 0.555 to
0.787 nats. The corresponding mean maximum assignment
probability changed from 0.848 to 0.793, from 0.771 to 0.706,
and from 0.788 to 0.688. Thus, $\nu=1$ produced softer masks on ambiguous bins. 

We compared raw HCA with backtracking on the first 16 test mixtures in every condition for both $\nu$ values. For each source-frequency component, the proposal $\mat{A}^{(n),\mathrm{HCA}}$ from $\mat{A}^{(n),\mathrm{old}}$ was tested along
$\mat{A}^{(n)}(2^{-k})=(1-2^{-k})\mat{A}^{(n),\mathrm{old}}+2^{-k}\mat{A}^{(n),\mathrm{HCA}}$, $k=0,\ldots,20$. The first trial whose exact matrix-dependent objective~\eqref{eq:em_Aterms} was at least the old value minus $10^{-9}$ was accepted. Otherwise, $\mat{A}^{(n),\mathrm{old}}$ was retained. A likelihood-decrease event was a drop exceeding $10^{-6}$ in frequency-wise average log-likelihood between EM iterations. Raw HCA produced 883,648 events, whereas backtracking produced none. Among the backtracking attempts, $k=0$ was accepted in 9.7\%, $1\leq k\leq20$ was accepted in 76.2\%, and the old matrix was retained in 14.1\%. Mean SDRi changed from 11.721 to $11.635\,\mathrm{dB}$ for $\nu=1$ and from 11.446 to $11.385\,\mathrm{dB}$ for $\nu=M$.

\subsection{Development-set $\nu$ sweep}
Figure~\ref{fig:sweep} varies $\nu$. For $d=-\vect{z}^{\Hh}\mat{A}\vect{z}\geq0$, the profile is $(1+2d/\nu)^{-(\nu+M)/2}$, and~\eqref{eq:u_mean} gives the EM data weight $(2/\nu)\bar u=(\nu+M)/(\nu+2d)$. 
For fixed $\nu$, this factor decreases with $d$, so observations
that fit the component less well receive smaller
weights in the scatter-matrix update. Relative to its value at $d=0$, the factor is reduced to $\nu/(\nu+2d)$ of that value, making this downweighting
stronger for smaller $\nu$.
At $\nu=M$, the cSTMM density reduces to the cACGMM density. As $\nu\to\infty$, the profile and weight tend to $\exp(-d)$ and 1, so the full and rank-1 models approach the cBMM and the cWMM, respectively. For $M=2$, the full and rank-1 parameterizations coincide, and their $(2,3)$ curves overlap. Full-cSTMM SDRi was maximized at $\nu=1$ for $(2,3)$, $(4,2)$, $(4,3)$, and $(4,4)$, and at $\nu=0.5$ for $(3,2)$ and $(3,3)$. Relative to $\nu=1$, $\nu=10^4$ reduced SDRi by 0.446--$6.399\,\mathrm{dB}$ and ambiguous-bin entropy by 0.192--$0.521\,\mathrm{nats}$ while increasing the mean maximum probability by 0.085--0.213. The large-$\nu$ limits therefore produced posterior masks closer
to one-hot assignments but lower SDRi than the best small-$\nu$
models on the development set.

\section{Conclusion}
The cSTMM places the cACGMM, the cBMM, and the cWMM within one directional mixture family controlled by $\nu$ and eigenvalue constraints. In this noise-free benchmark, the globally selected $\nu^\ast=1$ yielded consistent but modest SDRi gains over $\nu=M$. The model reduces to the cACGMM at $\nu=M$ and approaches the cBMM/cWMM for large $\nu$. Objective-checked backtracking removed the observed log-likelihood decreases but did not improve SDRi on the evaluated subset. Future work will address reference-free $\nu$ selection and evaluation with noise and downstream recognition.


\begin{thebibliography}{99}

\bibitem{araki2007}
S.~Araki, H.~Sawada, R.~Mukai, and S.~Makino,
``Underdetermined blind sparse source separation for arbitrarily arranged multiple sensors,''
\emph{Signal Process.}, vol.~87, no.~8, pp.~1833--1847, 2007.

\bibitem{mandel2010}
M.~I.~Mandel, R.~J.~Weiss, and D.~P.~W.~Ellis,
``Model-based expectation-maximization source separation and localization,''
\emph{IEEE Trans. Audio Speech Lang. Process.}, vol.~18, no.~2, pp.~382--394, 2010.

\bibitem{vuh10}
D.~H.~Tran Vu and R.~Haeb-Umbach,
``Blind speech separation employing directional statistics in an expectation maximization framework,''
in \emph{Proc. IEEE ICASSP}, 2010, pp.~241--244.

\bibitem{sawada2010perm}
H.~Sawada, S.~Araki, and S.~Makino,
``Underdetermined convolutive blind source separation via frequency bin-wise clustering and permutation alignment,''
\emph{IEEE Trans. Audio Speech Lang. Process.}, vol.~19, no.~3, pp.~516--527, 2011.

\bibitem{ito2016cBMM}
N.~Ito, S.~Araki, and T.~Nakatani,
``Modeling audio directional statistics using a complex Bingham mixture model for blind source extraction from diffuse noise,''
in \emph{Proc. IEEE ICASSP}, 2016, pp.~465--468.

\bibitem{ito2016cACGMM}
N.~Ito, S.~Araki, and T.~Nakatani,
``Complex angular central Gaussian mixture model for directional statistics in mask-based microphone array signal processing,''
in \emph{Proc. EUSIPCO}, 2016, pp.~1153--1157.

\bibitem{wang2018overview}
D.~Wang and J.~Chen,
``Supervised speech separation based on deep learning: An overview,''
\emph{IEEE/ACM Trans. Audio Speech Lang. Process.}, vol.~26, no.~10, pp.~1702--1726, 2018.

\bibitem{yoshioka2015}
T.~Yoshioka \emph{et al.},
``The NTT CHiME-3 system: Advances in speech enhancement and recognition for mobile multi-microphone devices,''
in \emph{Proc. IEEE ASRU}, 2015, pp.~436--443.

\bibitem{heymann2015}
J.~Heymann, L.~Drude, A.~Chinaev, and R.~Haeb-Umbach,
``BLSTM supported GEV beamformer front-end for the 3rd CHiME challenge,''
in \emph{Proc. IEEE ASRU}, 2015, pp.~444--451.

\bibitem{boeddecker2018}
C.~Boeddecker, J.~Heitkaemper, J.~Schmalenstroeer, L.~Drude, J.~Heymann, and R.~Haeb-Umbach,
``Front-end processing for the CHiME-5 dinner party scenario,''
in \emph{Proc. CHiME Workshop}, 2018, pp.~35--40.

\bibitem{raj2023gss}
D.~Raj, D.~Povey, and S.~Khudanpur,
``GPU-accelerated guided source separation for meeting transcription,''
in \emph{Proc. Interspeech}, 2023, pp.~3507--3511.

\bibitem{cornell2023chime7}
S.~Cornell \emph{et al.},
``The CHiME-7 DASR Challenge: Distant meeting transcription with multiple devices in diverse scenarios,''
in \emph{Proc. CHiME Workshop}, 2023, pp.~1--6.

\bibitem{mardia1999}
K.~V.~Mardia and I.~L.~Dryden,
``The complex Watson distribution and shape analysis,''
\emph{J. R. Stat. Soc. Ser. B}, vol.~61, no.~4, pp.~913--926, 1999.

\bibitem{kent1994}
J.~T.~Kent,
``The complex Bingham distribution and shape analysis,''
\emph{J. R. Stat. Soc. Ser. B}, vol.~56, no.~2, pp.~285--299, 1994.



\bibitem{kent1997}
J.~T.~Kent, ``Data analysis for shapes and images,''
\emph{Journal of Statistical Planning and Inference}, vol.~57,
no.~2, pp.~181--193, 1997.

\bibitem{kato2004}
S.~Kato and K.~Shimizu,
``A further study of $t$-distributions on spheres,''
Keio University Research Report KSTS/RR-04/012, 2004.



\bibitem{KatoMcCullagh2014}
S.~Kato and P.~McCullagh,
``A characterization of a Cauchy family on the complex space,''
arXiv:1402.1905, 2014.

\bibitem{panayotov2015}
V.~Panayotov, G.~Chen, D.~Povey, and S.~Khudanpur,
``LibriSpeech: An ASR corpus based on public domain audio books,''
in \emph{Proc. IEEE ICASSP}, 2015, pp.~5206--5210.

\bibitem{hadad2014mird}
E.~Hadad, F.~Heese, P.~Vary, and S.~Gannot,
``Multichannel audio database in various acoustic environments,''
in \emph{Proc. IWAENC}, 2014, pp.~313--317.


\end{thebibliography}
\end{document}